\documentclass[modern]{aastex63}

\submitjournal{ApJ}

\shorttitle{Hale Cycle Clock}
\shortauthors{Chapman, S.~C. et al.}

\graphicspath{{./}{figures/}}

\begin{document}

\title{The Sun's magnetic (Hale)  cycle and 27 day recurrences in the $aa$ geomagnetic index.}

\correspondingauthor{Sandra C Chapman}
\email{S.C.Chapman@warwick.ac.uk}

\author[0000-0003-0053-1584]{S. C. Chapman}
\affiliation{Centre for Fusion. Space and Astrophysics, Physics Department,
University of Warwick,
Coventry CV4 7AL, UK}
\author[0000-0002-7369-1776]{S. W. McIntosh}
\affiliation{National Center for Atmospheric Research, P.O. Box 3000, Boulder, CO~80307, USA.}
\author[0000-0002-6811-5862]{R. J. Leamon} 
\affiliation{University of Maryland--Baltimore County, Goddard Planetary Heliophysics Institute, Baltimore, MD 21250, USA.}
 \affiliation{NASA Goddard Space Flight Center, Code 672, Greenbelt, MD~20771, USA.}
\author[0000-0003-4484-6588]{N. W. Watkins}
\affiliation{Centre for Fusion. Space and Astrophysics, Physics Department,
University of Warwick,
Coventry CV4 7AL, UK}
\affiliation{Centre for the Analysis of Time Series, London School of Economics and Political Science, London, WC2A 2AZ, UK}
\affiliation{Faculty of Science, Technology, Engineering and Mathematics, The Open University,  Milton Keynes, MK7 6AA, UK}

\begin{abstract}

We construct a new solar cycle phase clock which maps each of the last 18 solar cycles onto a single normalized epoch for the approximately 22 year Hale (magnetic polarity) cycle, using the Hilbert transform of daily sunspot numbers (SSN) since 1818.
The occurrences of solar maxima show almost no Hale cycle dependence, confirming that the clock is synchronized to polarity reversals.  The odd cycle minima lead the even cycle minima by $\sim 1.1$ normalized years, whereas the odd cycle terminators \citep{McIntosh2019} lag the even cycle terminators  by $\sim 2.3$ normalized years.  The mimimum-terminator interval is thus relatively extended for odd cycles and shortened for even ones.
We re-engineer the \citet{SargentR27,SargentNEW}  R27 index and combine it with our epoch analysis to obtain a high time resolution parameter for 27 day recurrence in $aa$, $\langle acv(27) \rangle$. This reveals that the transition to recurrence, that is, to an ordered solar wind dominated by high speed streams, is fast, occurring within 2-3 solar rotations or less. It resolves an extended late declining phase which is approximately twice as long on even Schwabe cycles as odd.
Galactic Cosmic Ray flux rises in step with $\langle acv(27) \rangle$ but then stays high.
Our analysis also identifies a
slow timescale trend in SSN that simply tracks the Gleissberg cycle. We find that this trend is in phase with the slow timescale trend in the modulus of sunspot latitudes, and in antiphase with that of the R27 index.
\end{abstract}


\section{Introduction} \label{sec:intro}
Hale observed that the magnetic polarity of sunspots follows a roughly 22 year cycle \citep{Hale}. This Hale (magnetic polarity) cycle spans two (approximately 11-years long) Schwabe cycles of the sunspot number \citep{Schwabe1844}. In the century since Hale's discovery there have been many works exploring the importance of the Hale cycle in developing a coherent picture of the climatology of solar activity that is intrinsically tied to our star's magnetic field (see e.g. \citet{livingrev} and refs. therein). 

The fact that the solar magnetic field undergoes a magnetic polarity reversal each Schwabe cycle is intrinsically linked to the concept of an `extended solar cycle' \citep{CliverReview}. In this extended picture, solar magnetic activity coherently originates at high latitudes ($\sim 55^\circ$) and migrates equatorward over nearly two decades \citep[e.g.,][]{wilson,McIntoshP}. About a decade after  leaving high latitudes the pattern gives rise to the familiar `butterfly diagram' of sunspot latitudes \citep{Maunder04}. The pattern continues to move towards the equator where the oppositely polarised magnetic bands in each hemisphere eventually terminate \citep{Mcintosh2014a,McIntosh2019}. The Hale cycle is thus a fundamental mode of solar activity and is intrinsic to our understanding of the process that generates the magnetic field in the first place \-- the solar dynamo \citep[see, e.g.,][and refs. therein]{McIntoshP}. 

At a coarse level the range of phenomenology exhibited as solar (and associated) activity approximately repeats every Schwabe cycle. A wealth of observations now exist for solar  activity, but whilst they do suggest multi-Schwabe cycle, and thus potentially Hale cycle climatology (e.g. \citet{basu}) they extend over only a few Schwabe cycles. However, there are a number of phenomena which are observed over a sufficient number of cycles to show a pronounced magnetic polarity, or Hale Cycle, dependence \citep[see, e.g.,][]{livingrev}. As well as sunspot magnetic polarity, these include Galactic Cosmic Ray flux (GCR) \citep{Jopkii1,Jopkii2} and how solar coronal activity translates into the level of disorder present in the solar wind and its space weather impact. There are records demonstrating the imprint of solar activity on solar wind structure and geomagnetic activity at earth which reliably span a number of Hale Cycles. Indeed, \citet{SargentR27} (\citet{SargentNEW} has constructed a current version)   used the longest continuous record of geomagnetic activity, the $aa$ index \citep{aadefn}, to construct the `R27 index' of 27 day, solar rotation recurrence. This index peaks every Schwabe cycle but \citet{SargentR27} (see also \citet{Cliver}) identified a Hale cycle dependence in the shape of the peaks. 
  
This paper focuses on observed quantities that span multiple Hale cycles and brings two innovations to bear that can extract climatology on that scale. First, the \citet{SargentR27} prescription for the R27 index was an algorithmic one; we present a re-engineered version, built directly on the autocovariance of the $aa$ index which is nowadays much more readily computed. Second, we recently \citep{chapsunclock} showed that the Hilbert transform of sunspot number time series can be used to map the irregular Schwabe cycle onto a regular normalized timebase, or solar cycle `clock'. Once constructed, this solar cycle clock cleanly organises the cycle variation of solar flare statistics, the F10.7 index, and the measured geomagnetic activity at earth. Applying this technique to the daily sunspot number (SSN) we construct a Hale cycle clock and use a superposed epoch analysis \citep{Chree1913} to investigate the organisation of solar climatology on the Hale cycle.

In Section 2 we  revisit Sargent's R27 index. R27 is an index based on a measure of autocorrelation of the $aa$ index, however since the underlying autocorrelation or autocovariance is highly variable, smoothing is required in order to discern underlying trends in the time domain. In Section 3 we use the Hilbert transform of daily sunspot number to construct a Hale cycle clock. As discussed in Section 4, once the autocovariance of the $aa$ index is mapped onto the Hale cycle clock, we can perform an averaging across successive Hale cycles rather than smoothing over time.  This reveals that the transitions between disordered, and 27 day recurrent structure in the $aa$ index occur on a timescale that is less than or of order 2-3 solar rotations. The duration of this 27 day recurrent structure is almost twice as long during even cycles compared to odd ones, and we compare it with the Hale cycle variation of other multi-cycle observations. 

The superposed epoch analysis of the revised R27 allows us to revisit the pioneering work of \citep{Mayaud, L+S1981, L+S1989, S+L1989, L+S1991}. This series of papers culminated in the deduction that the $aa$ index is indicating that the solar cycle has two distinct components, (in the language of Legrand \& Simon, `dipole' and `toroidal' components), which are out of phase relative to one another \-- the `toroidal' component exhibiting a 5\--6 yr delay relative to the `dipole' component. In the discussion in Section 5  we consider their finding in the context of R27 and of the extended solar cycle. In this way we can start to unite the `light' (active regions, etc) and `dark' (coronal holes, solar wind, etc) sides of solar activity as natural and coupled manifestations of the Hale Cycle. Such efforts may also go a long way to revealing {\em why} the $aa$ index at solar minimum is an effective precursor of the upcoming maximum in sunspot number \citep{Feynman}.

Finally, identification and subtraction of a slow timescale trend in SSN is intrinsic to the Hilbert transform method (e.g. \citet{chapsunclock} and refs. therein). 
The slow timescale trend in SSN that forms part of our analysis (presented in Section 6) simply tracks the Gleissberg cycle \citep{gleiss}. We find that this trend is in phase with the slow timescale trend in the modulus of sunspot latitudes, and in antiphase with the R27.

\section{A re-engineered R27 index} \label{sec:style}
\citet{SargentR27} originally obtained the cross correlation coefficients between successive 27 day intervals of the $aa$ index \citep{aadefn,Mayaud} and then performed truncation and smoothing  to produce the original R27 index. We will base our analysis on the  autocovariance of the $aa$ index which is available from 1 January 1868. 
For a real-valued discrete signal $x_i$ the raw ($R_m$) and normalized ($acv(m)$) covariance \citep{Bendat} of a sequence with itself (i.e. the ``autocovariance") as a function of lag $m$ is, for $m \ge 0$:
\begin{eqnarray}
R_m&=&\sum_{n=0}^{W-m-1}\left(x_{n+m}-\frac{1}{W} \sum_{i=0}^{W-1}x_i \right)\left(x_n-\frac{1}{W} \sum_{i=0}^{W-1}x_i \right)\\
acv(m)&=&\frac{R_m}{R_0}
\end{eqnarray}
with the symmetry property that for $m < 0$, $acv(m)=acv(-m)$. In the above, the autocovariance is obtained for the sample window $i=1 \ldots W$ of the $x_i$. We will consider two measures of the level of ordered structure in the solar wind, inferred from $acv(27)$, the strength of the autocovariance of the $aa$ index at an $m=27$ day autocovariance lag.

The first of these is a `re-engineered' R27 index, hereafter `acv-R27', that performs the same analysis as that of \citet{SargentR27,SargentNEW} and in addition can be straightforwardly modelled and reproduced at each stage of its construction. Since, as we will demonstrate, the $acv(27)$ timeseries is highly variable, to extract any trend in it requires some averaging or smoothing in the time domain.  We calculate the autocovariance of the daily average of the $aa$ index at all lags $m$ for a $W$ day window centred on each day of the record. A running mean of length $S>W$ is then performed on the  $m=27$ day values, $acv(27)$ to obtain acv-R27. This procedure captures the essential operation of the original R27 index but has the advantage that since all lags are calculated, we can (i) test that $acv(27)$ is indeed a maximum of the autocovariance and (ii) use one of the other lags, say $acv(10)$ to establish a noise floor for the index. 
 
Our second measure will avoid smoothing in the time domain, desirable both to obtain higher time resolution and to avoid the possibility that smoothing can introduce  spurious periodicities in finite length, noisy timeseries \citep{Slutsky,Yule}. It will rely instead on averaging at the same Hilbert phase across successive Hale cycles and we will discuss it in more detail in section 4.1. 
 
We will first use model data to test the properties of the autocovariance $acv(27)$, and the acv-R27 index and will then compute them using the $aa$ index.  The model is of a positive definite sequence that incorporates (i) Poisson distributed noise added to a sporadically occurring spike signal which occurs either (ii) in short 27 day recurrent runs or (iii) randomly. The model time series is plotted in panel (a) of Figure 1, and intervals indicated with green bars are shown expanded in Figure 2. We use a  Poisson distribution with  probability mass function   $f_P(z\mid \lambda)=\frac{\lambda^{z}}{z!}e^{-\lambda}$, where $z$ is an integer, to generate random samples $z_{k,\lambda}$ that are both discrete, and positive definite, as is the $aa$ index. In the  model the noise is the sequence $5z_{k,1}$ with $\lambda=1$. The signal is the sequence  $z_{k,100}$ with $\lambda=100$. Each modelled `day' of data is then a randomly generated value for the noise, plus that of the signal if it has occurred on that day. Early in the model data sequence, the signal is organised into short runs of spikes with a 27 day recurrence as can be seen in panel (a) of Figure 2. These model recurrent co-rotating streams. The runs are randomly spaced, with a spacing $1+z_{k,200}$ and $\lambda=200$ and each run contains a random number of streams $1+z_{k,2}$, $\lambda=2$. The only non-random feature of these streams is their 27 day recurrence. For $t <1000$ the amplitude of the signal is twice that in the rest of the sequence. Next in the model data there is an interval of noise only, see panel (b) of Figure 2. Finally for $t>3000$ the signal is comprised of randomly occurring (uniformly distributed)  spikes, see panel (c) of Figure 2.

We now calculate $acv(27)$ and acv-R27 for this test data using an autocovariance window of $W=100$ and a running mean of length $S=1000$. Panel (a) of Figure 1 shows the test data. Panel (b) plots the running `daily' autocovariance for lag 27, $acv(27)$ (black) and lag 10, $acv(10)$ (red) and we see that $acv(27)$ discriminates the 27 day recurrent runs from the noise:  $acv(27)$ is well above the noise estimated as $acv(10)$. The `daily' $acv(27)$ is also insensitive to the amplitude of the recurrent runs. Panel (c) plots the daily $acv(27)$  as a colour map and we can see a peak at lag $m=27$ and more faintly at twice that value. Panel (d) plots $acv(27)$ (black) and its running mean (red) which is the acv-R27 index. We see that acv-R27 discriminates the 27 day recurrent runs with a smoothed value of $\sim 0.2$. The bottom panels plot examples of the individual $acv(27)$ for all lags $m$, and $m=27$ is indicated in grey, it is clearly a maximum of the autocovariance where the modelled recurrent streams are present. Where there is an absence of recurrent streams we do not obtain a peak in the autocovariance except at zero lag as we would expect for $\delta$ correlated random signals. This model data then confirms that acv-R27 has the essential properties of R27: it discriminates 27 day recurrences and is insensitive to amplitude. It has the additional advantage that we can directly obtain a noise floor from the data, and can check that the $m=27$ lag autocovariance $acv(27)$ is indeed the dominant $m \not = 0$ peak. 

\begin{figure}
\centering
\includegraphics[width=1.0\linewidth]{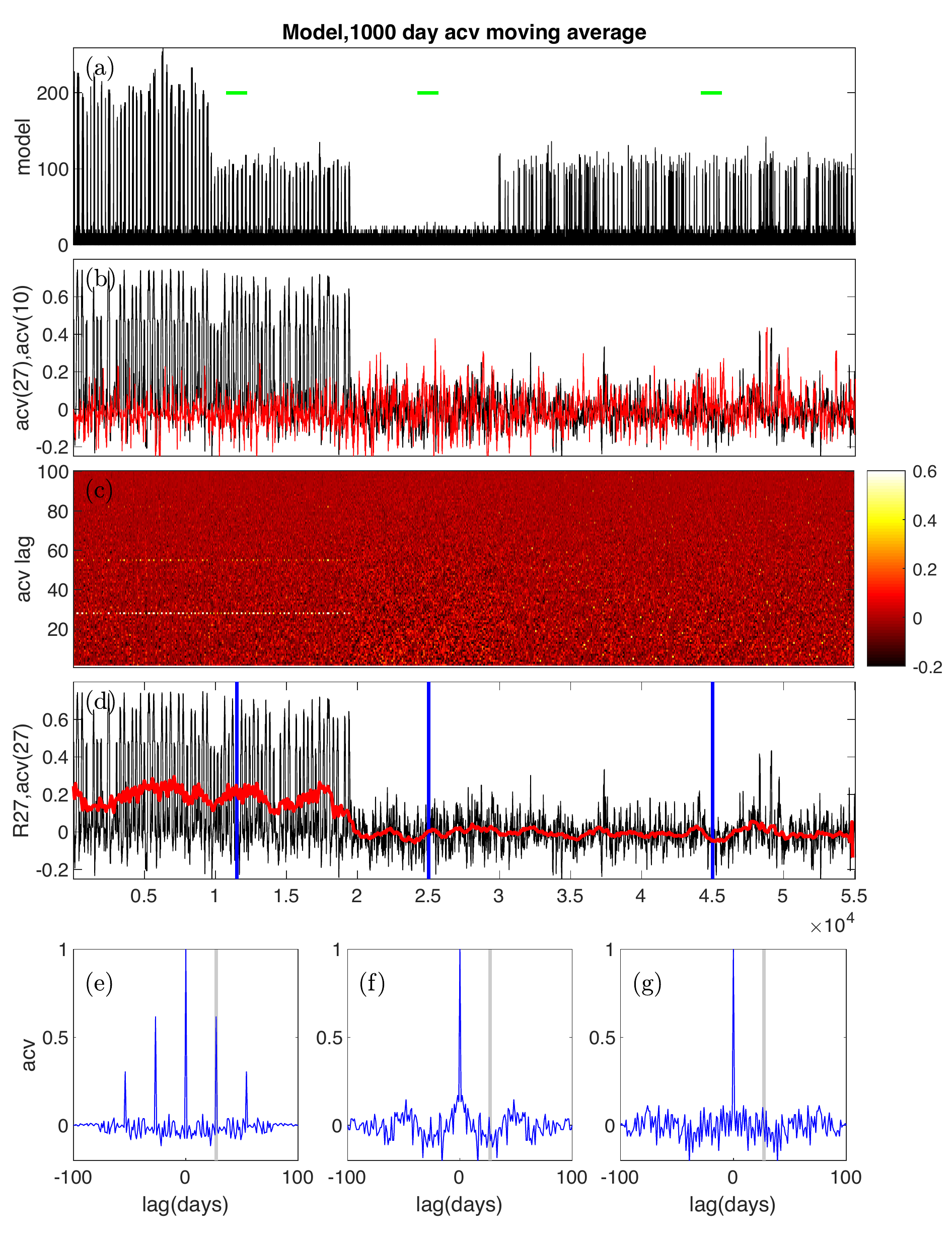}
\caption{Model data test of the acv-R27 method. The panels plot (a) the modelled data (b) the running 100 sample auto-covariance at lag 27 (black)  and at lag 10 (red) (c) the running sample covariance from lags m=5-100 as a colormap (d) the running 100 sample auto-covariance at lag 27 (black) and its 1000 point running mean (red), blue lines indicate the times for which samples (e-g) auto-covariance are plotted. Green bars indicate 1500 point intervals which are shown enlarged in Figure 2.}
\end{figure}

\begin{figure}
\includegraphics[width=1.0\linewidth]{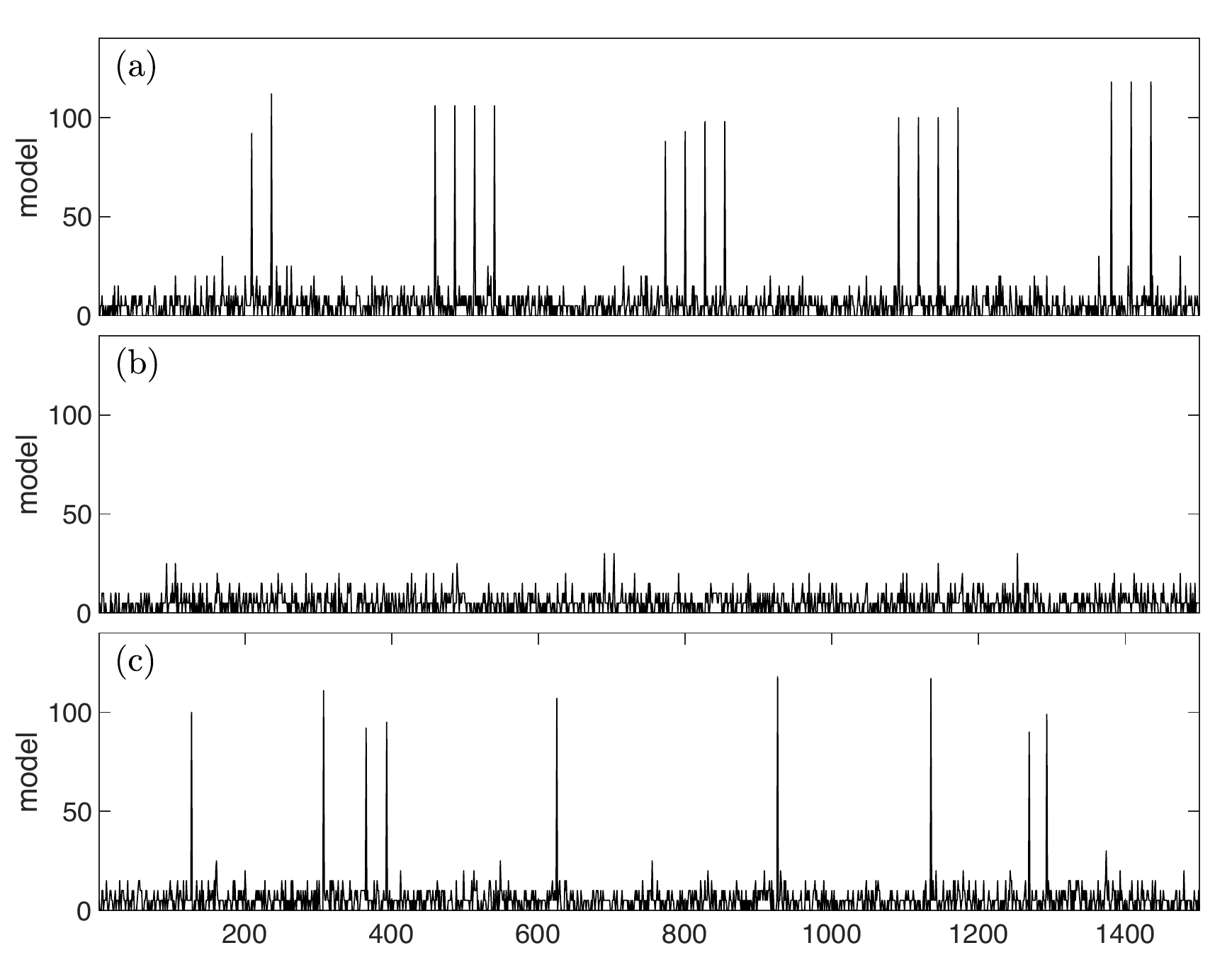}
\caption{Enlargements of the three 1500 point long intervals shown in Figure 1 (green bars).(a) 27 day recurrent signal plus noise (b) noise (c) randomly occurring signal plus noise.}
\end{figure}

We now use the same procedure to generate acv-R27 from the $aa$ index. This is shown in Figure 3.  We will start with daily $aa$ values obtained by averaging the 8 values of the 3 hourly $aa$ record for each day, shown in panel (a). Using the same parameters as above, we obtain daily samples of $acv(27)$ shown in panel (b), we can see 
that the  high values of the daily $acv(27)$ are significantly above that at $acv(10)$;
The daily $acv(m)$ are plotted at all lags in panel (c) and we can see that the autocovariance  has its main $m \not = 0$ peak at lag $m=27$, and  a weaker peak at $m=2 \times 27$. Panel (d) plots the daily and smoothed $acv(27)$ which is the acv-R27 index (analogous to R27). Individual daily $acv(m)$ are plotted (e,f) which again show high values of  acv-R27 correspond to peaks in the autocovariance at $m=27$ whereas when   acv-R27 takes a low value, there are no significant autocovariance peaks at $m \not = 0$. 

\begin{figure}
\centering
\includegraphics[width=1.0\linewidth]{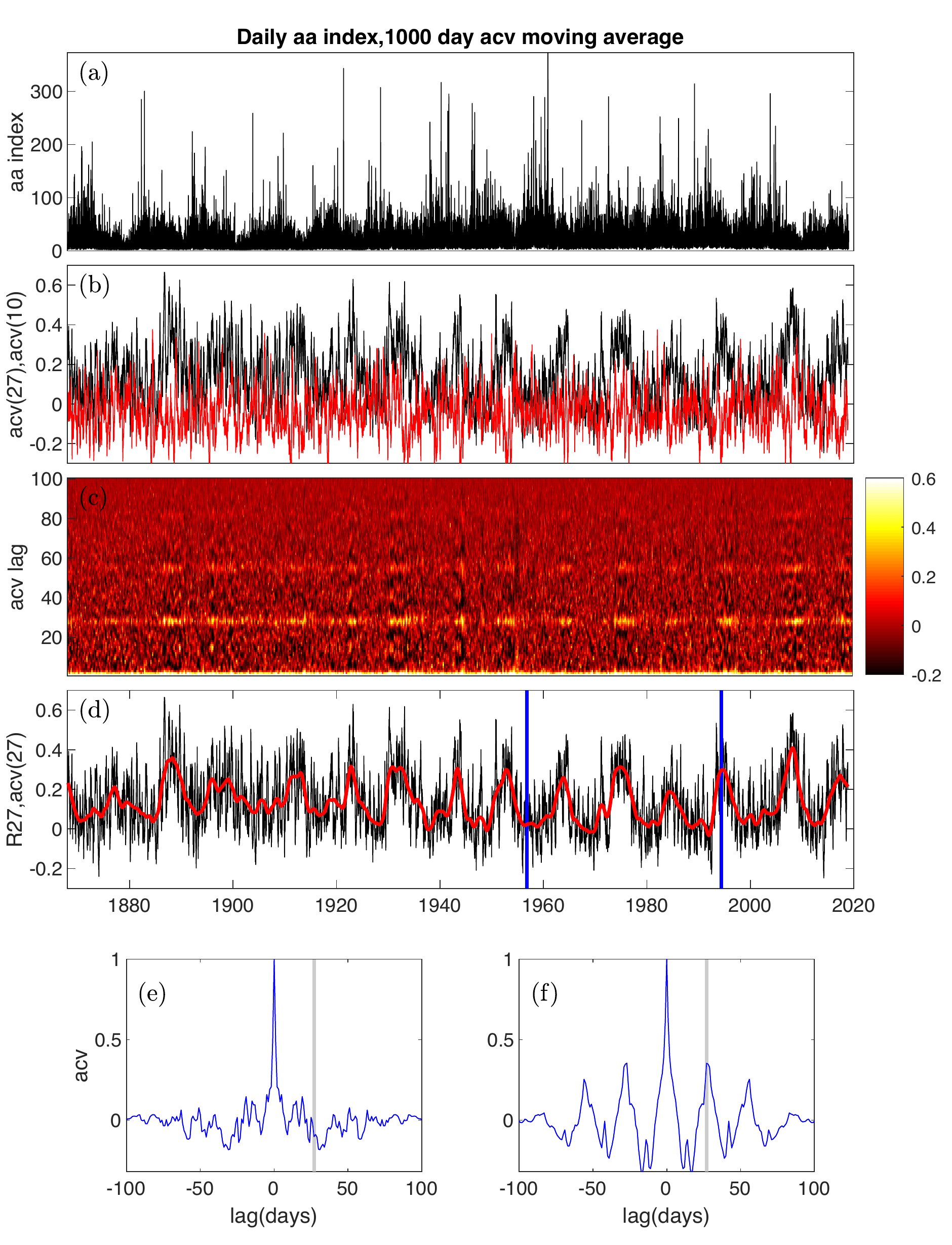}
\caption{The acv-R27 index based on the daily $aa$ index. From top to bottom (a) the daily $aa$ index (b) the daily autocovariance $acv(27)$ at lag 27 (black) and at lag 10 $acv(10)$ (red) (c) the running sample daily autocovariance from lags m=5-100 as a colormap (d) the daily $acv(27)$  (black) and its 1000 point running mean (red), the acv-R27, blue lines indicate times where individual daily autocovariance functions are plotted, for a time where the acv-R27 is relatively low (e) and high (f).}
\end{figure}

\section{Hilbert transform - building the sun clock}
\citet{chapsunclock} recently proposed a new method which orders  solar cycle variation on a timebase standardised to the solar cycle. The Hilbert transform of daily sunspot number (SSN) is used to map the irregular duration solar cycle in time onto a regular cycle in phase. We will use the same analysis as in \citep{chapsunclock} to obtain a phase-time mapping from the daily SSN record onto a standardized Hale cycle. This mapping can then be used (i) to study the Hale cycle climatology of the acv-R27 index and (ii) to perform an averaging across multiple standardized Hale cycles of $acv(27)$, giving a higher time resolution measure of the level of 27 day recurrent structuring of the solar wind, which we can then compare with the Hale climatology of other quantities available across multiple cycles.

\begin{figure}
\centering
\includegraphics[width=0.9\textwidth]{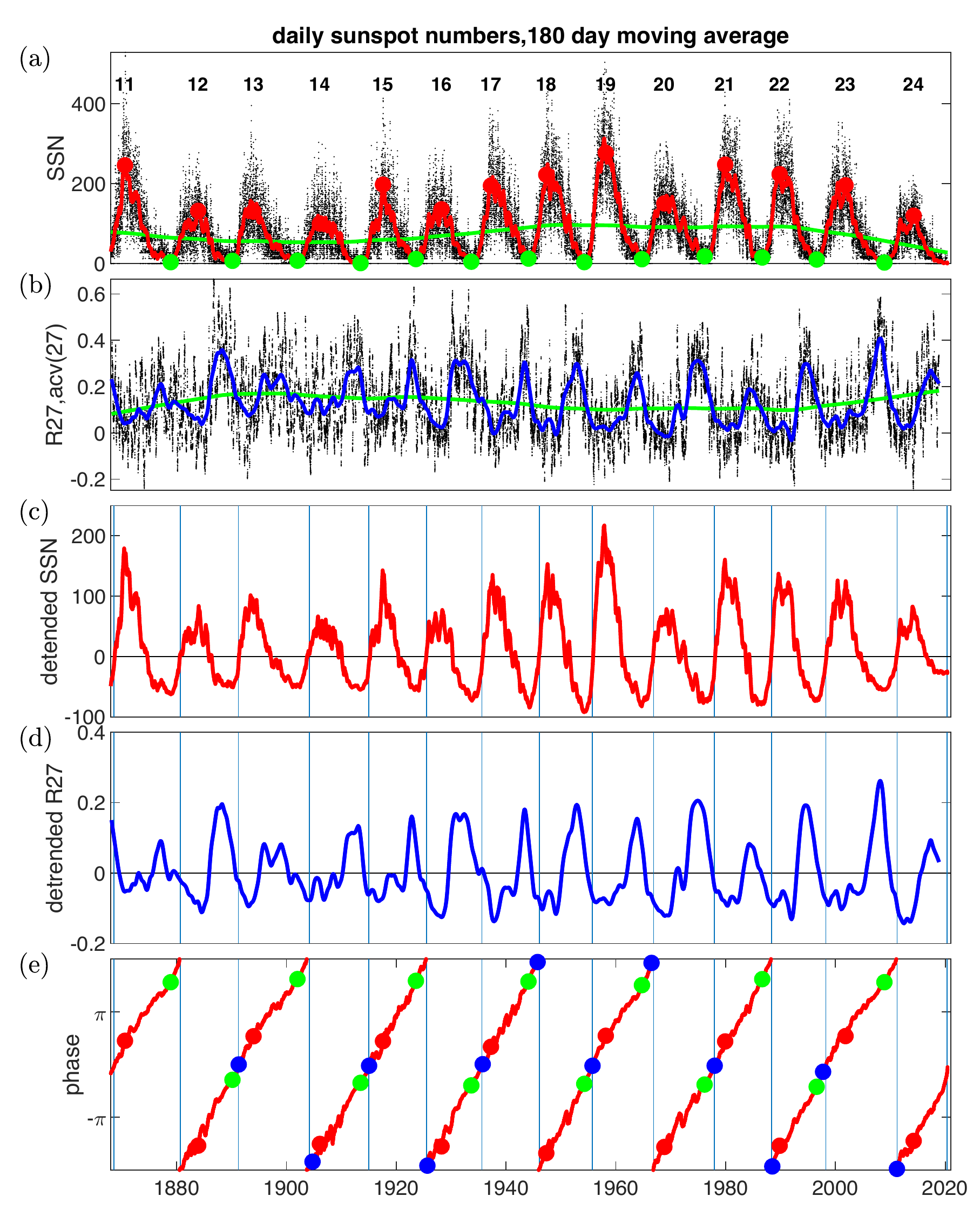}
\caption{From top to bottom (a) daily sunspot number (black), 180 day moving average (red) and slow timescale trend obtained by local regression  on a 40 year window (green); (b) daily $acv(27)$ values (black), their 1000 day moving average acv-R27 (blue) and the slow timescale trend obtained by local regression  on a 40 year window (green); (c) smoothed sunspot number with slow timescale trend subtracted (d)  acv-R27 with slow timescale trend subtracted (e) analytic phase of smoothed detrended sunspot number (red) plotted on a $[-2\pi,  2 \pi]$ interval. The maxima and minima are indicated by red and green circles respectively and the blue circles indicate terminators from \citet{McIntosh2019}. Zero phase is set to be at the average phase of the terminators. Solid  blue vertical lines indicate terminator times of occurrence (see text).}
\end{figure}

We use the daily SSN record which provides an almost uninterrupted measure of solar coronal activity since 1st January 1818. It is plotted from the start of the $aa$ index record (1 January 1868) in Figure 4(a).   We can decompose this time series  $S(t)$  in terms of a time-varying amplitude $A(t)$ and phase $\phi(t)$ by obtaining its analytic signal \citep{Gabor1946,Boashash1992} $A(t) exp[i\phi(t)]$  such that the real part of this signal is $S(t)$ and the imaginary part is obtained such that $A(t) exp[i \phi(t)]=S(t) +i H(t)$ where $H(t)$ is the Hilbert transform of $S(t)$. This provides a mapping between time and signal phase, that converts the (variable) duration of each solar cycle into a corresponding uniform phase interval. \citet{chapsunclock} considered a mapping of each Schwabe cycle to phase in the range $0$ to $2\pi$. Here we will use this same mapping but will consider the Hale cycle which corresponds to the phase range $0$ to $4 \pi$, that is, two successive Schwabe sunspot cycles.

While defined for an arbitrary time series, the analytic signal  only gives a physically meaningful decomposition of the original time series if  the instantaneous frequency  $\omega(t) =d\phi(t)/dt$  remains positive \citep{Boashash1992}. We therefore need to remove fast fluctuations and, for a positive-definite signal such as the daily sunspot number, a background trend. Following \citet{chapsunclock} before performing the Hilbert transform  we  performed a $T_s=180$ day moving average. We obtained a slow timescale trend by performing a robust local linear regression which down-weights outliers (`rlowess') using a $T_B=40$ year window. We subtract the slow timescale trend (green line in Figure 4(a)) to give a sunspot time series that is unambiguously zero-crossing (Figure 4(c)). We then obtain the Hilbert transform $H(t)$ for this  smoothed and detrended signal which then gives the analytic signal. We obtain the slow timescale trend for acv-R27 in exactly the same manner. These slow timescale trends are also an aspect of solar climatology which we will consider in the context of the Gleissberg cycle in section 6. The acv-R27 index is plotted in 4(d) with this slow timescale trend subtracted. Panel (e) plots the Hilbert analytic phase obtained from the daily SSN record, this is wrapped to a domain of $[ -2 \pi,  + 2 \pi]$ Hale cycle. Overplotted are the maxima (red circles), minima (green circles) and terminators (blue) of each Schwabe cycle. Zero ($\pm 2 \pi)$ phase is set to the average 
phase of the terminators and 
vertical blue lines are drawn at the times when the analytic phase crosses zero and $\pm 2 \pi$
to demarcate one Schwabe cycle from the next. Panel (c) shows the same suggestion of a Hale cycle climatology in acv-R27 as that originally identified by \citet{SargentR27} in R27, alternate Schwabe cycles have longer/shorter peaks in acv-R27. Comparing panels (c) and (d) we also find a new result-   that the downward sweeps in acv-R27 correspond to the observed terminator times identified in \citet{McIntosh2019}. Termination of each Schwabe cycle has a corresponding switch-on of disorder in the solar wind. However as discussed above, acv-R27, and indeed the original R27 index, are by necessity smoothed in time. We will overcome this limitation by mapping the observations over each cycle onto a uniform `clock'. Whilst the cycle lengths are irregular, panel (e) shows how they can be mapped to a regular interval in phase. One can either construct a mapping based on the Schwabe cycle as in \citet{chapsunclock}, or as we will do here, construct a mapping based on successive pairs of Schwabe cycles, that is, on Hale cycles, which corresponds to a mapping between irregular cycles in time, and a regular $4 \pi$ interval in analytic phase. 

\section{Hale Cycles \-- The 22-Year Sunclock}
We now use the mapping between time and phase plotted in Figure 4(e) to construct plots of solar climatology on a regular, normalized Hale cycle. An example of the resulting Hale cycle clock is shown in Figure 5, which overplots successive Hale cycles on a $[0 , 2 \pi]$ interval corresponding to 22 normalized years. This is mapped from $[0 , 4\pi]$ in Hilbert analytic phase of the daily SSN record, that is, two successive Schwabe cycles.
 
\begin{figure}
\includegraphics[width=\linewidth]{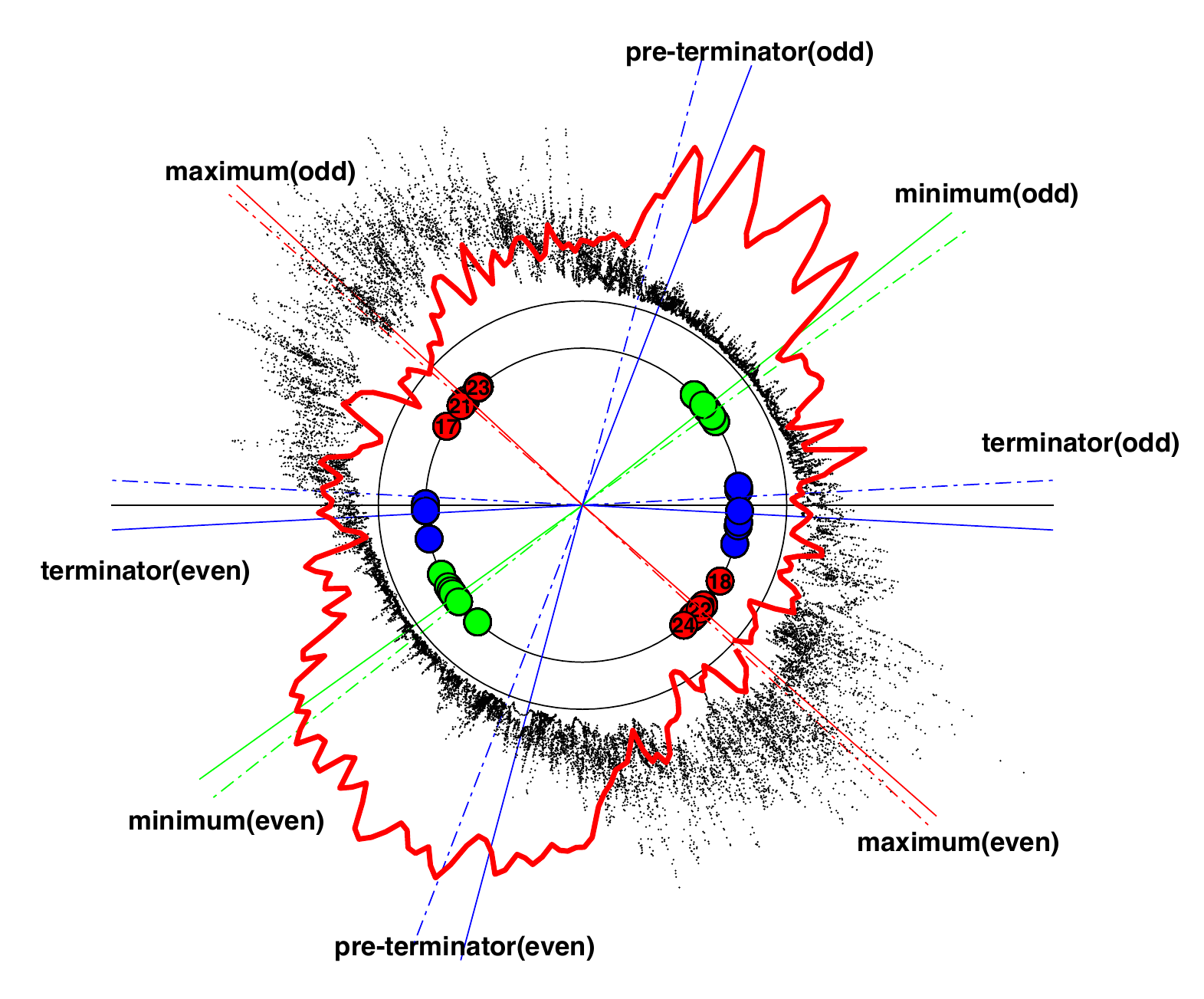}
\caption{Hale cycle clock constructed such that increasing time (analytic phase) is read clockwise. $[0, 2\pi]$ in phase on the clock corresponds to the normalized 22 year Hale cycle (phase $[0, 4 \pi ]$ from the Hilbert transform of daily sunspot number). 
The analytic phases of the maxima and minima of the last 18 solar cycles are indicated by red and green circles respectively and the blue circles indicate terminators for the last 12 solar cycles \citep{McIntosh2019}. Red, green and blue solid lines indicate the average analytic phases for the maxima, minima and  respectively; these averages are performed separately for odd and even Schwabe cycles. The pre-terminators (blue line) are at phase $2 \pi/5$ (4.4 normalized years) in advance of the terminator. The solid lines indicating these average phases are continued across the clock with dash-dot lines- these indicate that whilst the maxima occur regularly, there is a Hale cycle variation in when the minima and terminators occur. The horizontal black line is located at the terminator occurrence averaged over all Schwabe cycles. These features form the basis of the Hale cycle clock. Black dots overplot daily F10.7. The red line is the $\langle acv(27) \rangle$ obtained by dividing the normalized 22 year Hale cycle into $12 \times 22$ month-long intervals and averaging the values of $acv(27)$ that fall within each of these intervals across multiple Hale cycles.}
\end{figure}
 
The sunspot maxima and minima along with the terminator occurrences from the previous figure are plotted on the clock. The averages of their locations are indicated by solid lines, these averages are obtained independently over odd and even Schwabe cycles. The solid lines are extended as dashed lines across to the opposite half of the clock, so that the dashed lines indicate where the average maxima, minima and terminators would occur if they repeated exactly from one Schwabe cycle to the next. This is indeed the case for the maxima; the even cycle maxima average is very close to $\pi$ in phase in advance of the odd cycle maxima average, they differ by $\sim 0.03$ radians or, for a normalized 22 year Hale cycle, $\sim 0.6$ normalized years. The polarity reversal of the solar polar field is known to occur close to the solar cycle maxima \citep{Thomas2014} so that the clock does indeed provide a robust epoch analysis tool for the Hale cycle. The odd cycle minima lead the even cycle minima by $\sim 0.05$ radians, or $\sim 1.1$ normalized years, whereas the odd cycle terminators lag the even cycle terminators by $\sim 0.11$ radians, or $\sim 2.3$ normalized years. The minimum-terminator interval is thus relatively extended for odd cycles and shortened for even ones. The daily F10.7 index since 14th February 1947 \citep{Ften} is overplotted (black dots) and we can see that as identified in \citet{LeamonH,chapsunclock}, the terminators coincide with an increase in activity seen in F10.7 as the activity of the next cycle commences \citep{Mcintosh2014a,McIntosh2019}. Each Schwabe cycle `switch-on' of activity at the terminators is preceded by a `switch-off' of activity, and \citet{chapsunclock} identified this pre-terminator as being approximately 4.4 normalized years in advance of the terminator which on this 22 year Hale cycle clock is $2 \pi/5$ in phase.  We emphasise that this is only an estimate of where the pre-terminator should be located, based upon observed correspondence with changes in multiple measures of solar activity \citep{chapsunclock,LeamonP}; such as low values of the F10.7 index as can be seen in Figure 5.
 
\subsection{Hale cycle structure of the $aa$ index}
\citet{SargentR27} noted that enhancements in the R27 index indicating recurrent behaviour are longer(shorter) on alternate Schwabe cycles, which we can also see in acv-R27 in Figure 4. We can now use the Hale cycle clock to perform a direct average of $acv(27)$ which avoids smoothing in the time domain, tracking the variation in 27 day recurrences at higher time resolution. We divide the normalized 22 year Hale cycle into $12 \times 22$ month-long intervals and average the values of $acv(27)$ that fall within each of these intervals over multiple Hale cycles. This is then detrended by subtracting the same (40 year window,  rlowess) slow timescale trend as for acv-R27. The resulting $\langle acv(27) \rangle$ will have a time resolution dictated by the time window over which the autocovariance is calculated, which is 100 days here. The $\langle acv(27) \rangle$ is plotted on Figure 5 (red line) and it reveals a clear Hale cycle dependence of 27 day recurrence in the $aa$ index. The $\langle acv(27) \rangle$ peaks during quiet intervals of solar activity as seen in F10.7 but the interval over which $\langle acv(27) \rangle$ is enhanced is almost twice as long during even cycles compared to odd ones. The $\langle acv(27) \rangle$ sharply increases around (odd cycles) or before (even cycles) the pre-terminator and  decreases at minimum, decaying to its low value at the terminator; the timescales for these sharp changes determined here are at the time resolution of the autocovariance window, that is about 3-4 solar rotations. The physical timescale of these changes could be faster than this.

As established in our modelling in section 2 (compare panels (a)  (b) and (d) of Figure 1), both $\langle acv(27) \rangle$  and acv-R27 are constructed to be sensitive to the time structure, i.e. recurrences, in the signal whilst being independent of the signal amplitude. Thus any Hale cycle variation does not simply reflect the overall level of activity in the $aa$ index. Furthermore, the $aa$ index (units, $nT$) is discretized in amplitude \citep{bub,ChapmanCar} since the underlying $K$ index \citep{Kindex} is a quasi-logarithmic 0-9 integer scale that characterizes the maximum  magnetic deviations that occur during each 3 hour period at a given observatory. Therefore we focus on the occurrences of $aa$ exceeding a threshold value, rather than the   absolute values of $aa$, or its time averages.

Figure 6 plots $\langle acv(27) \rangle$ on a Hale cycle clock as in Figure 5, and black dots are plotted at successively increasing radii to indicate days where the daily maximum $aa$ index exceeded thresholds $T_{aa}=100,200,300,400,500,600nT$. Extreme space weather events are seen as radial 'spokes' on this plot. The most extreme events occur in a disordered sequence with relatively low values of $\langle acv(27) \rangle$. However, more moderate $aa \lesssim 200nT$ events do occur when $\langle acv(27) \rangle$ takes a high value. The occurrence rates are plotted in Figure 7, which shows histograms of counts of days within non-overlapping normalized 6 month bins in which the $aa$ index exceeds $T_{aa}$. Panel (a) of Figure 7 plots $\langle acv(27) \rangle$ and vertical blue lines are drawn at the upcrossings of 
$\langle acv(27) \rangle =0$. The vertical blue lines indicate the transition from an $aa$ index signal that is disordered, to one that contains 27 day recurrences, and $\langle acv(27) \rangle$ rises steeply at these transitions. It is well known that geomagnetic storms are preferentially triggered by irregularly occurring  coronal mass ejections (CMEs) around solar maximum, whereas they are likely to be  triggered by corotating interaction regions (CIRs) in the declining phase \citep{Richardsonrev,pulrev}; the CIR driven events are more moderate than the CME driven ones \citep{alv,bor}. Our analysis is consistent with this picture and in addition reveals that the transition between these two behaviours is fast; from the time interval just following solar maximum where   $\langle acv(27) \rangle$ is at its low value, to a late-declining phase (shaded yellow in Figure 7)  where $\langle acv(27) \rangle$ rises steeply to its high value and geomagnetic storms decrease in their intensity. The frequency of occurrence of more moderate events does not decrease until around solar minimum. The duration of the late-declining phase shows a Hale cycle dependence, it is approximately twice as long for even cycles as for odd ones.

\begin{figure}
\includegraphics[width=\linewidth]{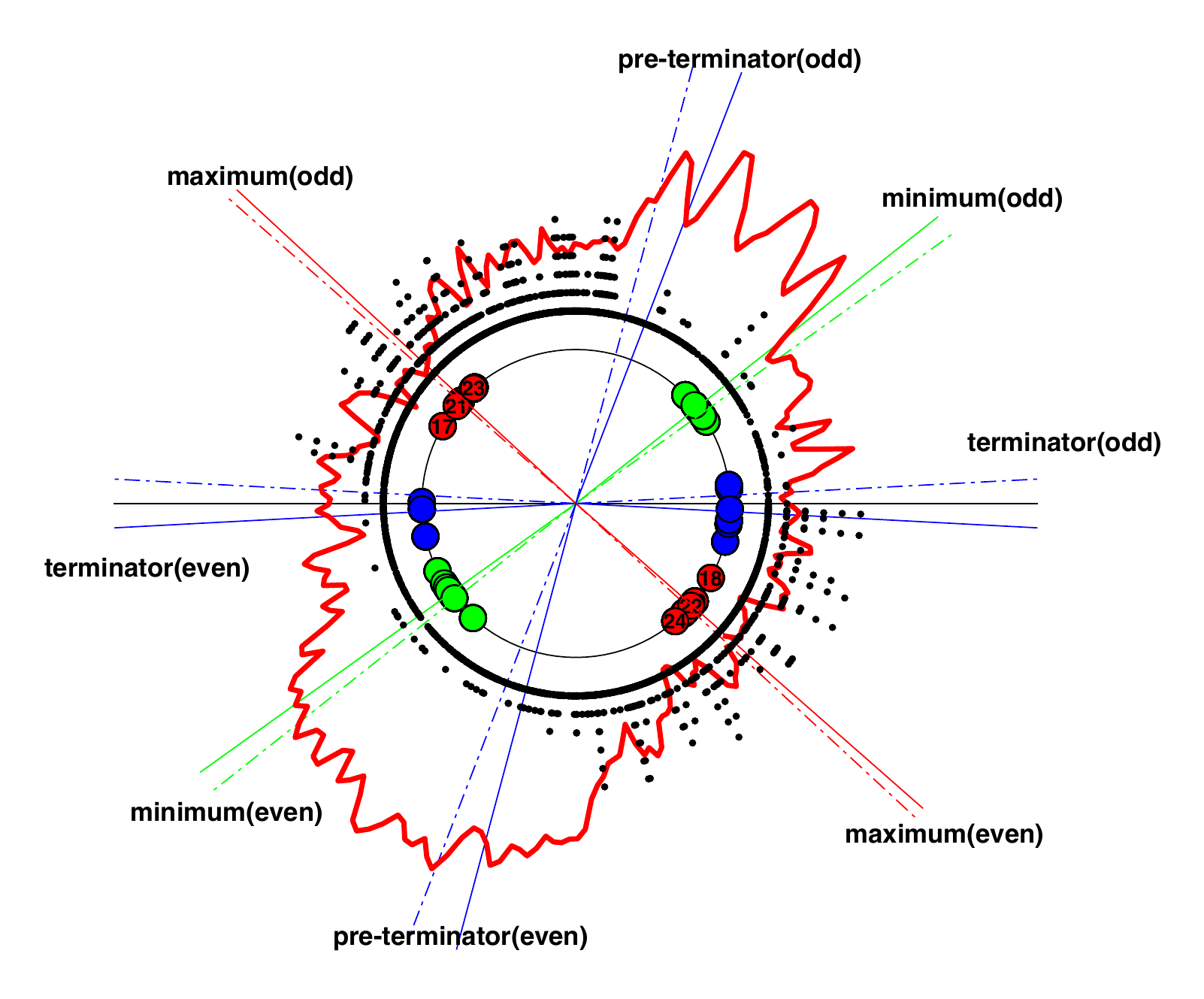}
\caption{Hale cycle clock constructed such that increasing time (analytic phase) is read clockwise, $[0 - 2\pi]$ in phase on the clock corresponds to the normalized 22 year Hale cycle (phase $[0 - 4 \pi ]$ from the Hilbert transform of daily sunspot number). The format is as the previous figure except that activity in the $aa$ index is over-plotted on the $\langle acv(27) \rangle$.
Black dots plotted at successively increasing radii indicate days where the daily maximum $aa$ index exceeded $100,200,300,400,500,600nT$.}
\end{figure}

\begin{figure}
\centering
\includegraphics[width=0.75\linewidth]{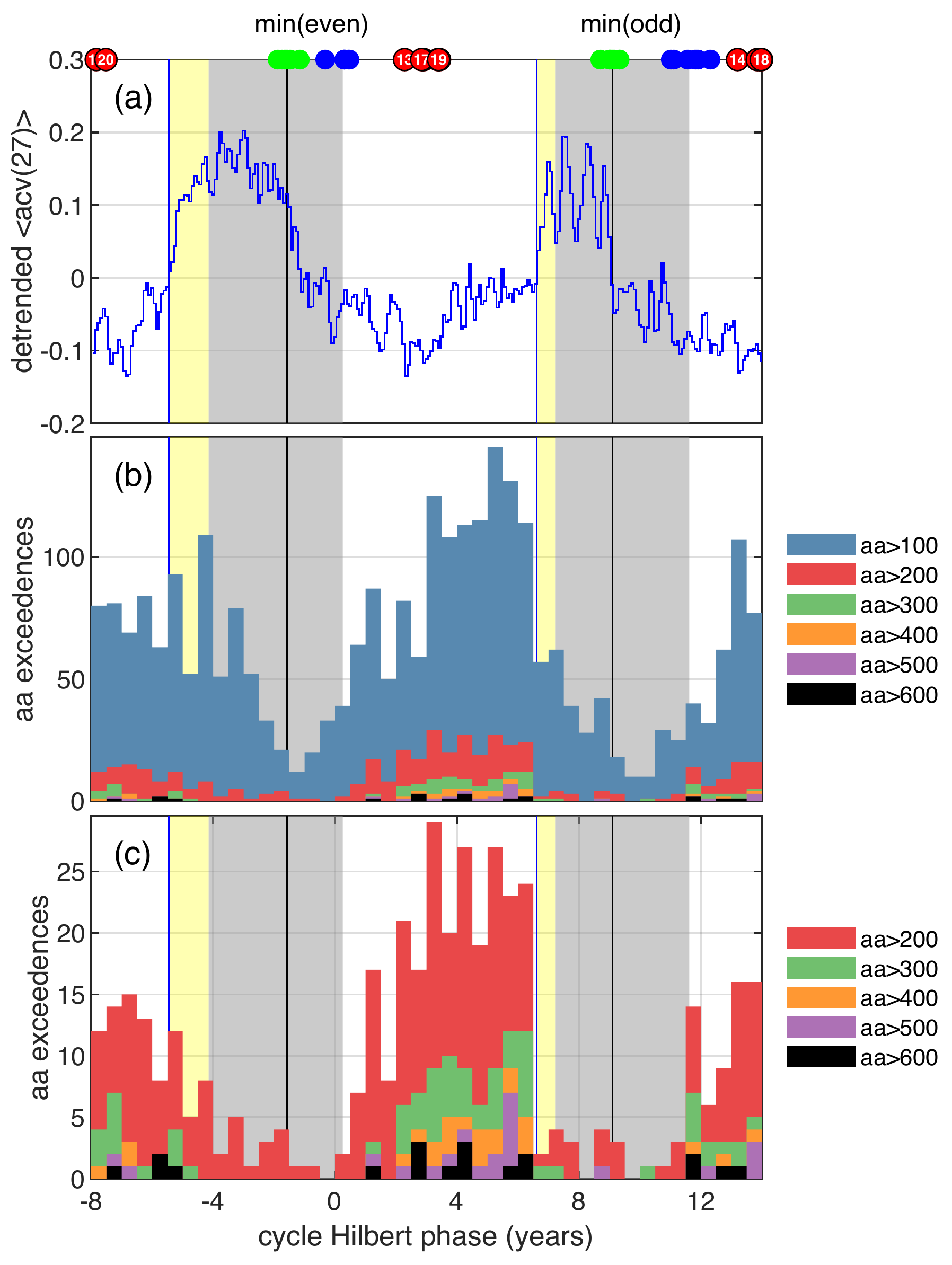}
\caption{ Hale cycle clock phase plotted linearly, the normalized 22 year Hale cycle (phase $[0- 4 \pi ]$ from the Hilbert transform of daily SSN).
The analytic phases of the maxima and minima of the last 18 solar cycles are indicated by red and green circles respectively and the blue circles indicate terminators for the last 12 solar cycles \citep{McIntosh2019}. Black lines indicate the average analytic phase for the minima. The quiet interval around minimum, shaded grey, is of 
 4.4 normalized years duration starting at the pre-terminator and ending at the average phase of the terminators.
 These averages are performed separately for odd and even cycles.  The late-declining phase (see text) is shaded yellow. Panel (a) $\langle acv(27) \rangle$ (blue trace) obtained by dividing the normalized 22 year Hale cycle into $12 \times 22$ month-long intervals and averaging the values of $acv(27)$ that fall within each of these intervals over multiple Hale cycles. Blue vertical lines are at the zero upcrossings; panel (b) 6 (normalized) month binned counts of days where the maximum daily $aa$ index value exceeded $100,200,300,400,500,600nT$ and panel (c) counts where the maximum daily $aa$ index value exceeded $200,300,400,500,600nT$. Zero and 11 normalized years on this plot correspond to the phase indicated by the black line in Figure 5.}
\end{figure}

\subsection{Hale cycle structure in GCR flux}
Galactic Cosmic Ray (GCR) flux is also known to show a 22 year pattern, with even cycles showing a longer, `flat top' enhancement compared to odd cycles. This has been attributed to different particle drift patterns when the northern solar pole has predominantly positive or negative  polarities \citep{Jopkii1,Jopkii2,Smith1,Smith2,Fer}. During positive polarities, cosmic ray protons reach Earth after approaching the poles of the Sun in the inner heliosphere and moving out along the heliospheric current sheet (HCS). During negative polarities, cosmic ray protons approach the Sun along the HCS plane and leave via the poles. Heliospheric modulation, both of the HCS and of the structure of the solar wind, has been found to play a significant role particularly in the declining phase of the solar cycle (\citet{Thomas2014} and refs. therein). Plots of annual mean SSN versus the annual mean GCR intensity are typically used to characterize the Hale cycle variation of GCR flux (see e.g. \citep{Ross} and refs. therein). Here, we can directly track GCR flux across the normalized Hale cycle. Since $\langle acv(27) \rangle$ is a measure of recurrence in the $aa$ index, it directly depends on recurrent structure in the solar wind, so that we can infer that high values of $\langle acv(27) \rangle$ correspond to a solar wind dominated by 27 day recurrent high speed streams, whereas low values of $\langle acv(27) \rangle$ correspond to a solar wind that is disordered. 

Figures 8 and 9 plot the daily GCR flux (since 30th April 1964, see \citet{oulu}) following subtraction of its slow timescale (40 year window rlowess) trend, and $\langle acv(27) \rangle$ on the Hale cycle clock. In Figure 8 we can see that the rise in GCR flux and $\langle acv(27) \rangle$ roughly track each other in the interval between the sunspot maximum and minimum, that is, through the declining phase. The GCR flux departs from the $\langle acv(27) \rangle$ at sunspot minimum, where $\langle acv(27) \rangle$ has passed its maximum value. The first half of the interval of enhanced GCR flux thus coincides with a solar wind dominated by recurrent streams, whereas the second half does not. 

Figure 9 shows that that the increase in GCR flux is  aligned with that in $\langle acv(27) \rangle$, particularly in even cycles.  The increase in GCR flux during the declining phase thus coincides with a solar wind dominated by recurrent streams, and as suggested by \citet{Thomas2014} this may dominate over the  polarity effect on proton drift paths during the declining phase. The comparison with $\langle acv(27) \rangle$ on a normalized 22 year cycles shows how clearly demarcated these different intervals are. If we read off the length of the interval of enhanced GCR flux as times when the detrended GCR flux is above zero in Figure 9, we can see that in even cycles it has approximately twice the duration as in odd cycles. The longer duration of the late-declining phase in even cycles identified in Figure 7 corresponds to $\sim 1.3$ normalized years of enhanced GCR flux during even cycles. The GCR flux in even cycles only begins to decline after the terminator, crossing zero in Figure 7 at $\sim 1.6$ normalized years after the terminator has occurred. Thus the extended duration 'flat top' of GCR flux seen in even cycles compared to odd ones may be partially but not fully explained by the extended duration of the interval of enhanced recurrent structure dominating the solar wind as seen in high values of $\langle acv(27) \rangle$. 
 
 \begin{figure}
\includegraphics[width=\linewidth]{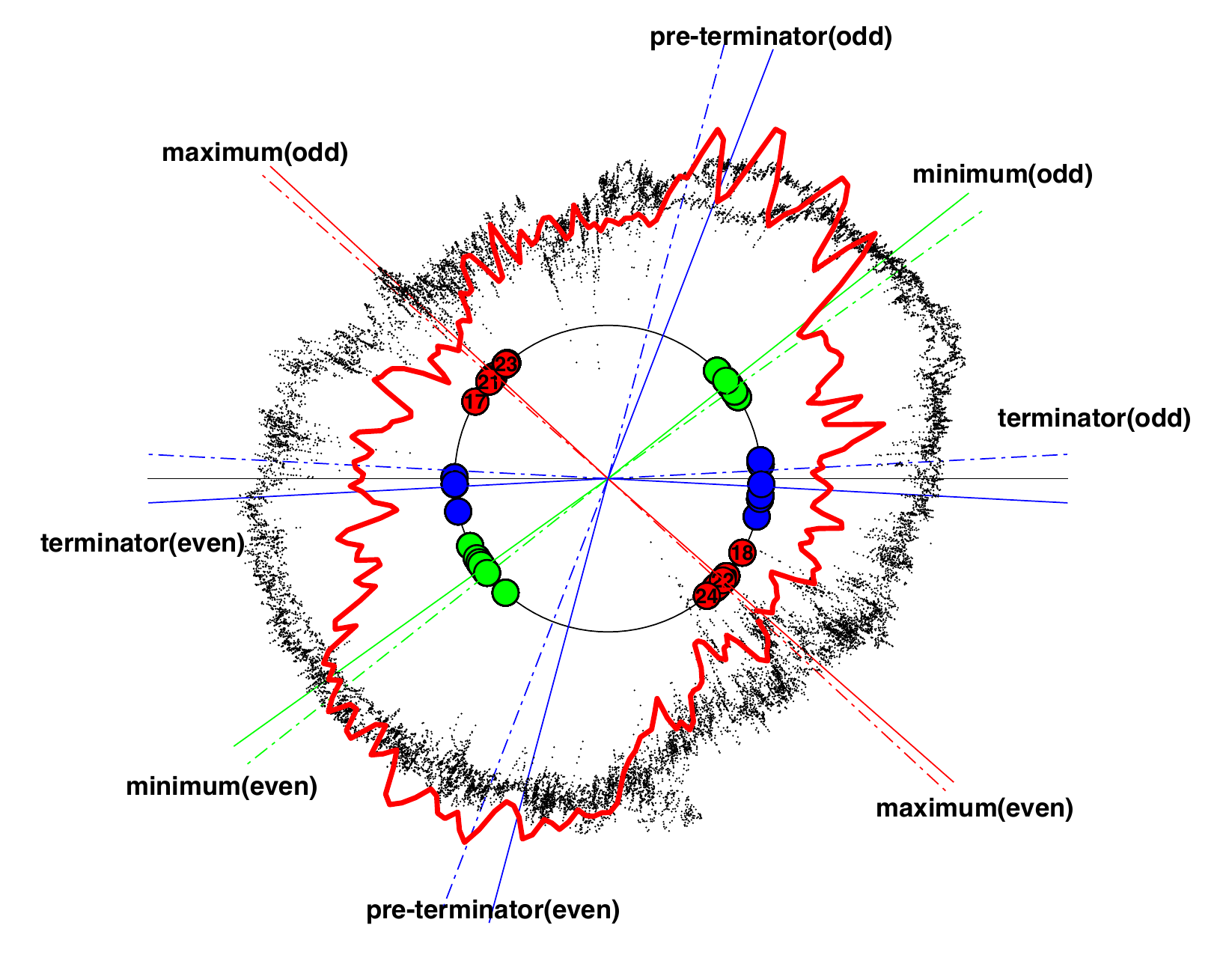}
\caption{ Hale cycle clock constructed such that increasing time (analytic phase) is read clockwise. $[0, 2\pi]$ in phase on the clock corresponds to the normalized 22 year Hale cycle (phase $[0, 4 \pi ]$ from the Hilbert transform of daily sunspot number). The $\langle acv(27) \rangle$  (red) is overplotted on the daily detrended cosmic ray flux (black).}
\end{figure}

\begin{figure}
\centering
\includegraphics[width=0.75\linewidth]{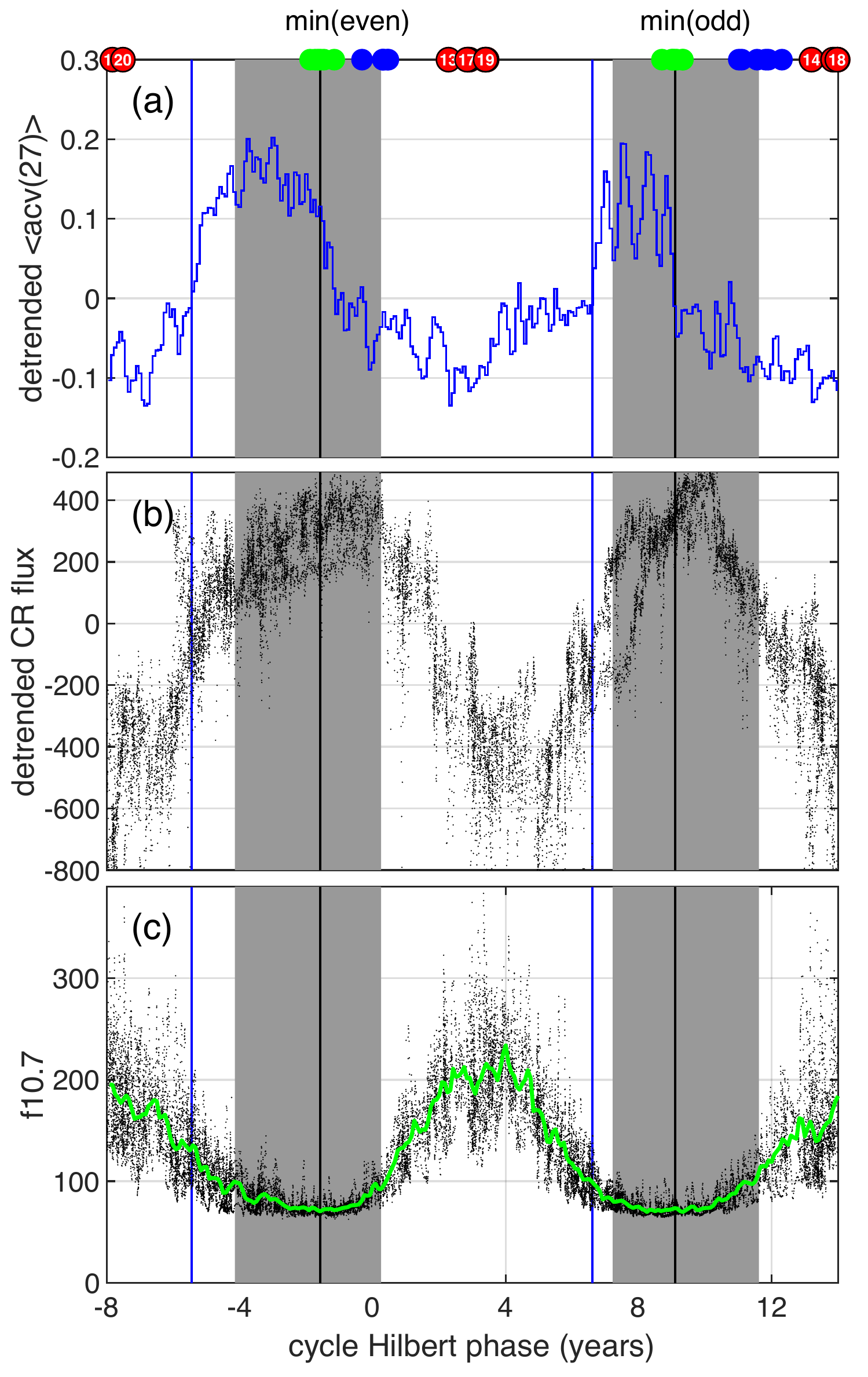}
\caption{Hale cycle clock phase plotted linearly, the normalized 22 year Hale cycle (phase $[0, 4 \pi ]$ from the Hilbert transform of daily SSN). The format follows that of Figure 7. Panel (a) $\langle acv(27) \rangle$ (blue trace) with blue vertical lines at the zero upcrossings; panel (b) detrended daily cosmic ray flux (black); panel (c) daily F10.7 (black) and its average (green) obtained by dividing the normalized 22 year Hale cycle into $12 \times 22$ month-long intervals and averaging the values of faily F10.7 that fall within each of these intervals over multiple Hale cycles.}
\end{figure}

\subsection{Hale cycle and sunspot latitudes}
The Greenwich record of sunspot latitudes and areas (1 May 1874 -  30 September 2016) provides a record of sunspot activity extending over multiple Hale cycles which we directly compare with $\langle acv(27) \rangle$ in Figure 10. On this figure, panel (a) plots $\langle acv(27) \rangle$ as in previous figures. The modulus of the latitudes of sunspot  centroids recorded each day are averaged, and the  sequence of these daily averages are plotted in panel (b) of Figure 10 (black dots). We obtained $\langle acv(27) \rangle$ by dividing the normalized 22 year Hale cycle into $12 \times 22$ month-long intervals and averaging the values of $acv(27)$ that fall within each of these intervals over multiple Hale cycles. We now perform the same operation on the daily averages of sunspot latitudes and this is plotted as the green trace on panel (b) of Figure (10) (we will denote this average sunspot latitude trace as $\langle SSL \rangle$). In \citep{Owens}  the average modulus sunspot latitude was proposed as  a useful parameterization of the Schwabe cycle as it shows a sudden increase at solar minimum due to the emergence of high latitude sunspot pairs of new cycle polarity. The same overall sawtooth pattern identified by \citet{Owens} can be seen here in $\langle SSL \rangle$, it up-crosses latitude $\sim 15^\circ$ at the  minima. However the $\langle SSL \rangle$ also shows a Hale cycle dependence, the sawtooth pattern is more sharply defined for even cycles than for odd ones, and whilst both the pre-terminators correspond to  $\langle SSL \rangle$ down-crossing  latitude $\sim 10^\circ$, the terminators occur when $\langle SSL \rangle$ is at a higher latitude $\sim 22^\circ$ for even cycles compared to odd ones $\sim 20^\circ$. Panel (c) plots the sunspot latitudes as a  classic butterfly diagram. Large sunspot areas are  not seen in the quiet interval, as we would expect, the largest switch off at the start of the declining phase identified from $\langle acv(27) \rangle$ (blue vertical lines). This is consistent with a relatively sharp transition from a disordered solar wind, populated with CMEs which result in large $aa>200$ geomagnetic storms, to a more ordered solar wind where storms, driven by recurrent high speed streams, are more moderate.
   
\begin{figure}
\centering
\includegraphics[width=0.75\linewidth]{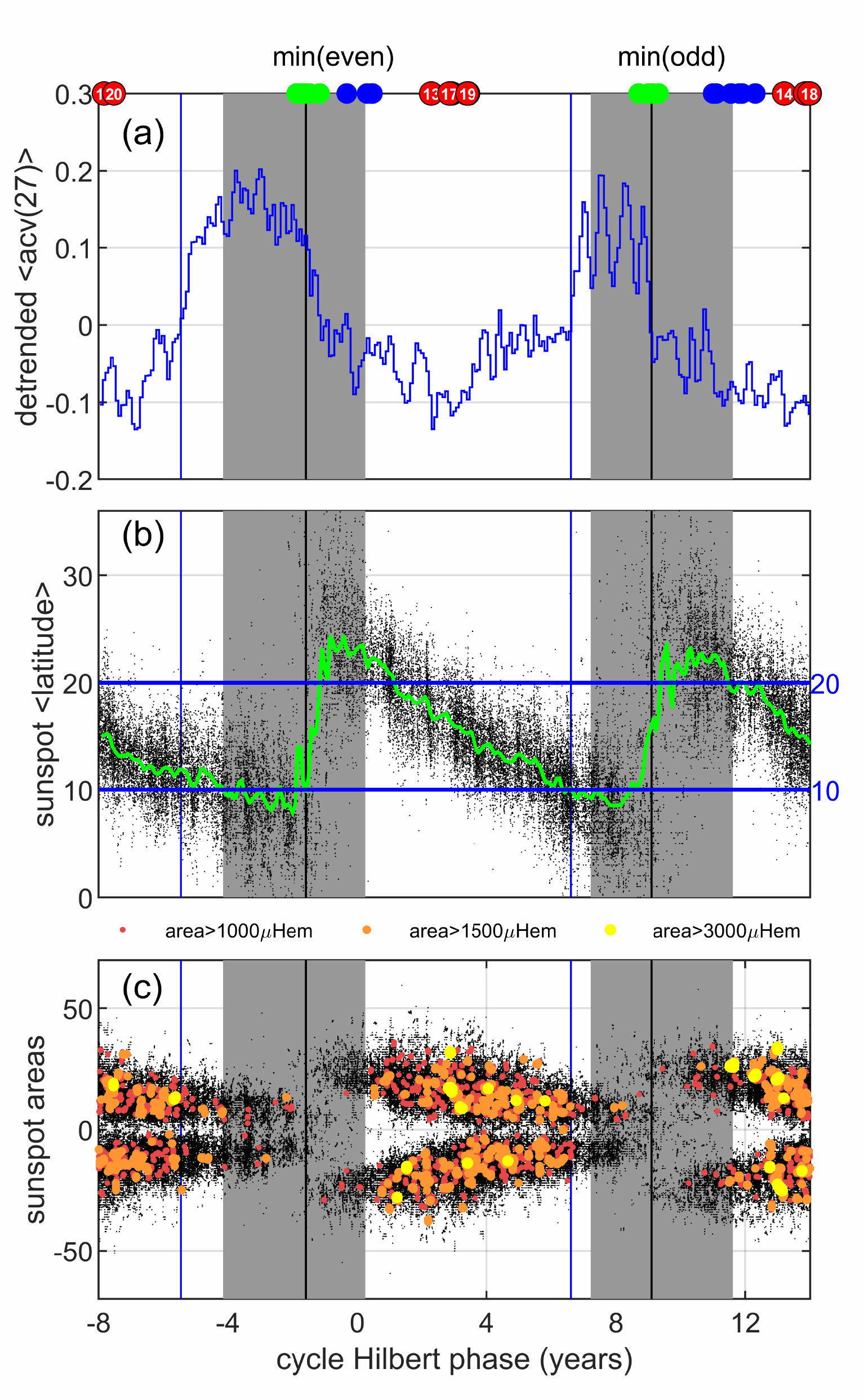}
\caption{ Hale cycle clock phase plotted linearly, the normalized 22 year Hale cycle (phase $[0, 4 \pi ]$ from the Hilbert transform of daily sunspot number). The format follows that of the previous figure. Panel (a) $\langle acv(27) \rangle$ (blue trace) with blue vertical lines at the zero upcrossings; panel (b)  the daily averaged modulus of  sunspot area centroid latitudes (black) and overplotted, its multi-Hale cycle variation (green) obtained by dividing the normalized 22 year Hale cycle into $12 \times 22$ month-long intervals and averaging the (modulus) sunspot area latitudes  that fall within each of these intervals over multiple Hale cycles. Panel (c) daily latitude positions of  the sunspot area centroids, symbol colours and sizes differentiate areas $>1000,1500,3000 \mu Hem$.}
\end{figure}

\section{Mayaud, Legrand, Simon and the Hale/Extended Cycle}
The analysis provided above allows us to briefly revisit, and contextualise, the pioneering work of \citet{Mayaud, L+S1981, L+S1989, S+L1989, L+S1991}, (see also \citet{Feynman}). This series of papers culminated in the deduction that the $aa$ index is indicating that the solar cycle has two distinct components which are out of phase relative to one another, one associated with active Sun and the other with quiescent (recurrent) high speed solar wind streams. Their hypothesis was that the bimodal behaviour of the $aa$ index was consistent with the Sun exhibiting a `two-component cycle.' The earlier component in their view, the `dipole' component occurs at mid\--to\--high latitudes and is exhibited some 5-6 years before the second, the `bipolar' component, that originates in the sunspot bearing latitudes before repeating.

Figure 11 overlays the
 trend-removed timeseries of the SSN and $acv-R27$ from Figure 4 and we can see that they are in antiphase.
 The acv-R27 rises to its peak  at or just after the  pre-terminator (regardless of its shape, see Figures 8, and 9). This is the epoch of long-lived mid-latitude coronal holes \citep{Krista, Hewins}. Those coronal holes are the sources of Mayaud, Legrand and Simon's dipole component recurrent high speed streams and belong to the magnetic band of the extended solar cycle that becomes host to the sunspots that follow some 5 or 6 years later \citep[see  Fig.~11C, or][]{McIntoshP}. Our analysis, when taken in concert, validates the insightful work of \citep{Mayaud, L+S1981, L+S1989, S+L1989, L+S1991}. 

 Going a step further, the temporal phasing of the dipolar and bipolar component signatures and the resulting correspondence with the magnetic systems of the Hale cycle provides insight into why the aa-index around sunspot minimum is a stronger than average precursor of the {\em upcoming} sunspot cycle strength \citep[see, e.g.,][]{Feynman}. The pre-terminator recurrent high-speed solar wind streams originate from the {\em same\/} Hale cycle magnetic bands as the sunspots that follow only a few years later \-- tying the `dark' and `light' sides of the Hale cycle together and illustrating the {\em why} of the strong precursor relationship, although the {\em how} requires further work.  

\begin{figure}
\centering
\includegraphics[width=1.0\linewidth]{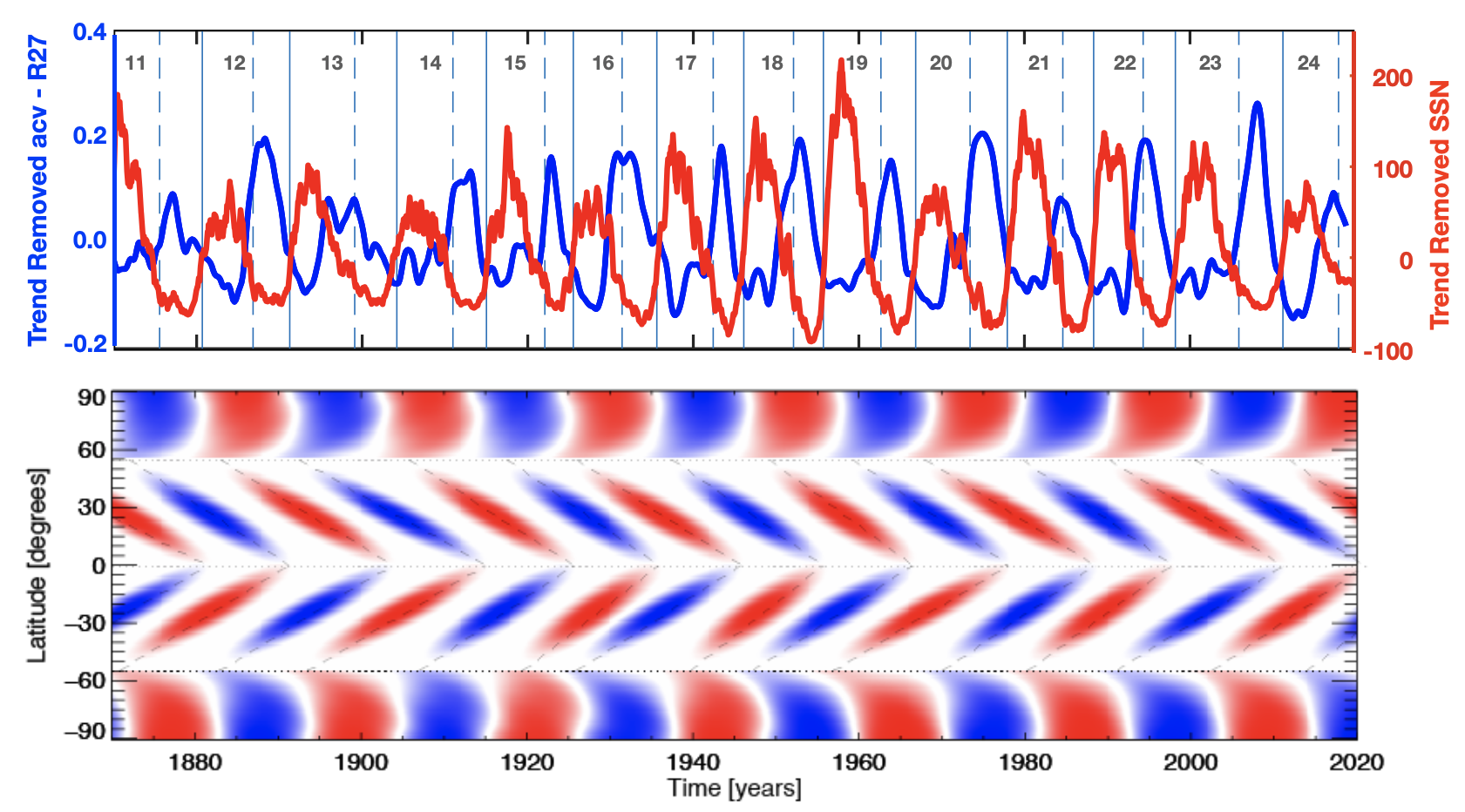}
\caption{Placing the (trend-subtracted) sunspot number in context with the (trend-subtracted) acv-R27 and the ``Extended Solar Cycle'' as a representation of the Hale Cycle.
The upper panel overlays 
 the smoothed sunspot number with slow timescale trend subtracted (red) and acv-R27 with slow timescale trend subtracted (blue). The lower panel plots the data-motivated Hale Cycle ``band\--o\--gram'' of \citep{Mcintosh2014a,McIntosh2019,McIntoshP}. The pre-terminators and terminators are indicated by vertical dashed and solid blue lines respectively.}
\end{figure}

\section{Slow timescale trends and the Gleissberg cycle}
In the above analysis we subtracted a slow timescale trend obtained by a robust local linear regression which down-weights outliers (`rlowess') using a $T_B=40$ year window from the SSN, acv-R27 (Figure 4) and GCR flux (Figures 8 and 9). As part of its construction, the $\langle acv(27) \rangle$ has the same trend subtracted as acv-R27. We did not subtract a slow timescale trend from the average sunspot latitude (Figure 10) as it is only $\sim 2^\circ$ in amplitude however we will consider it here. These slow timescale trends are compared to the Gleissberg cycle in Figure 12. The Gleissberg cycle $G^{(N)}$ is defined \citep{gleiss} in terms of sucessive solar maxima as:
\begin{equation}
G^{(N)}=\frac{1}{8}\left( R_{max}^{(N-2)}+2R_{max}^{(N-1)}+2R_{max}^{(N)}+2R_{max}^{(N+1)}+R_{max}^{(N+2)}  \right)
    \end{equation}
where $R_{max}^{(N)}$ is the SSN at the maximum of the $Nth$ solar cycle. We used values for $R_{max}^{(N)}$ determined by SILSO.  The $G^{(N)}$ are plotted on panels (a-c) of Figure 11 (blue lines-symbols). Panel (a) overplots the SSN trend which can be seen to track the Gleissberg cycle so that the slow timescale trend that we have removed in order to obtain the Hilbert transform of SSN is just that of the  Gleissberg cycle. Panel (b) shows that there is a slow timescale trend in acv-R27 that is approximately in antiphase to the Gleissberg cycle. The average sunspot latitude (panel (c)) trend is in phase the Gleissberg cycle. 

This may relate to the slow timescale trend in reconstructed IMF and open solar flux which is relatively low during 1890-1910 and relatively high during 1960-1990 (\citet{Lockwood2011,Owens2012}, see also \citet{Sval}). Those reconstructions \citep{Lockwood2011} rely in part on the magnitude of the $aa$ index. The acv-R27 and $\langle acv(27) \rangle$ as we have shown with the modelling in section 2, are not sensitive to variation  in the overall amplitude of $aa$, instead they capture its time structure which Figure 12 would then suggest that enhanced 27 day recurrence coincides with reduced open solar flux. 
The $aa$ index is known to contain a systematic drift in the offsets used in its construction, as discussed in detail in \citet{lockwoodnewaa1}. Since acv-R27 and $\langle acv(27) \rangle$ are not sensitive to variation in the overall amplitude of $aa$, we would not expect this systematic drift to appear in the slow timescale trend shown in Figure 12. We have verified that this is indeed the case by repeating our analysis with the homogenised $aa$ that has been corrected for these systematic effects by \citet{lockwoodnewaa2}, our results are unchanged.

\begin{figure}
\centering
\includegraphics[width=0.75\linewidth]{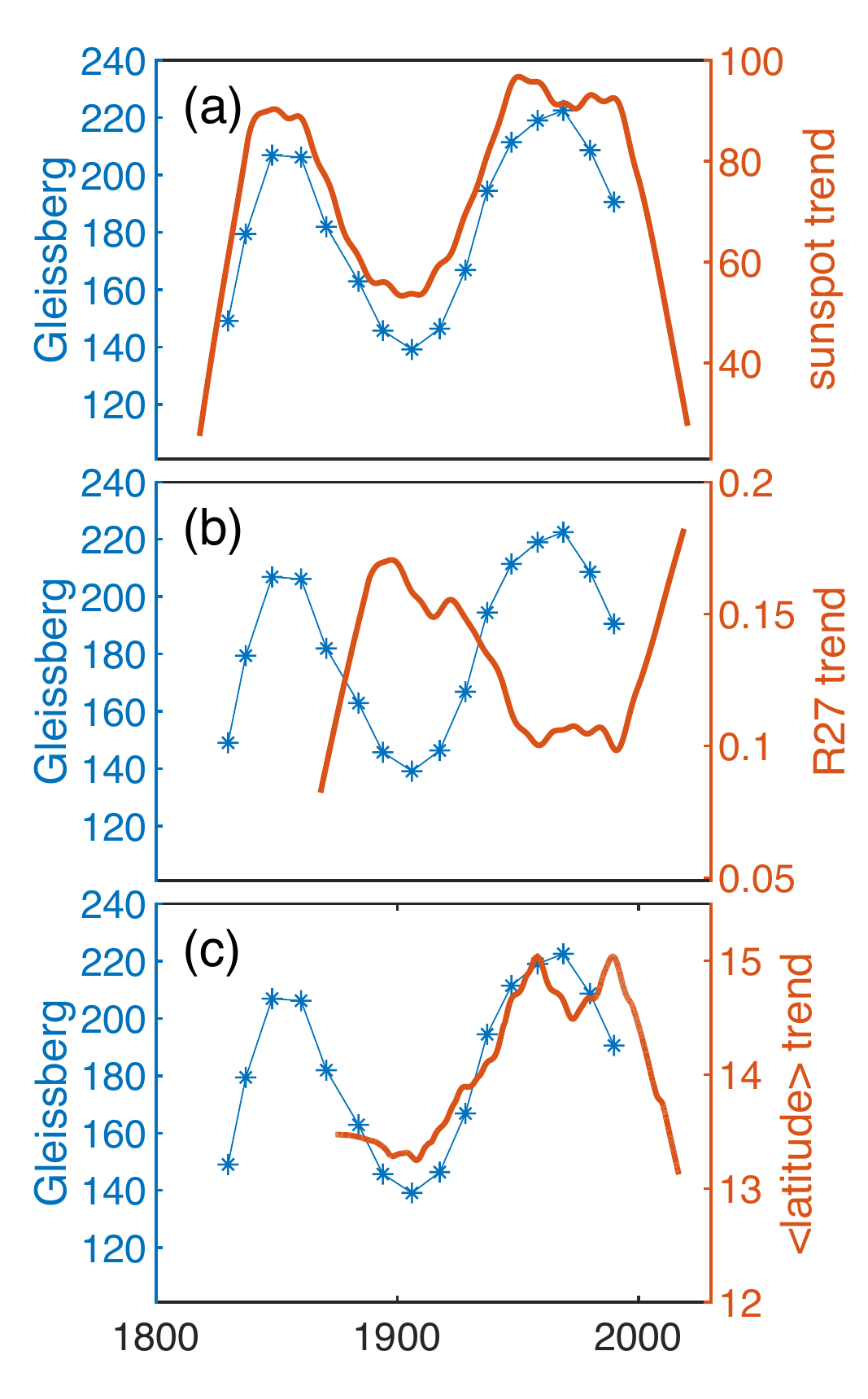}
\caption{Gleissberg cycle overlaid on slow timescale (40 year rlowess local linear regression) trends of (a) SSN, (b) acv-R27 and (c) average modulus sunspot latitude.}
\end{figure}

\section{Conclusions}
We have used the Hilbert transform of daily sunspot number (SSN) to construct a Hale cycle clock which affords the comparison of quantities observed over multiple solar cycles. This clock normalizes successive Hale cycles onto a regular 22 (normalized) year timebase. The autocovariance of the $aa$ index at a lag of 27 days has been calculated directly and then averaged across these multiple normalized Hale cycles. This provides an index ($\langle acv(27) \rangle$) of solar rotation recurrences with time resolution at the autocovariance window length (here, 100 days)
rather than being smoothed in time.

We obtained the following results:
\begin{itemize}

\item The $\langle acv(27) \rangle$ can resolve changes on a time resolution of 2-3 solar rotations and it changes on (or less than) this timescale.

\item The $\langle acv(27) \rangle$ rises sharply at the declining phase which is extended  on even cycles---twice as long as odd.


\item The $\langle acv(27) \rangle$ drops sharply, on the timescale of a few solar rotations, at solar minimum then slowly decays to its `low' value at the terminators.

\item The occurrence of the solar maxima shows almost no Hale cycle dependence (so that the Hilbert transform derived clock is indeed locked to polarity reversals which occur close to maxima) whereas the minima and terminators do show a Hale cycle dependence. The odd cycle minima lead the even cycle minima by   $\sim~1.1$ normalized years, whereas the odd cycle terminators lag the even cycle terminators by $\sim 2.3$ normalized years. The mimimum-terminator interval is thus relatively extended for odd cycles and shortened for even ones. 

\item The GCR flux rises in step with $\langle acv(27) \rangle$ that is, at the onset of a more ordered solar wind, but then stays high.

\item Average sunspot latitude shows a Hale cycle dependence.

\item The fact that there are Hale cycle dependencies on the parameters that we have explored would imply that there are previously unnoticed dependencies on the Sun's magnetic configuration. Pointing to the lower panel of Figure 11 for assistance, for odd numbered cycles (for example cycle 15) the toroidal band configuration is -/+/-/+ North to South while it is +/-/+/- North to South for even numbered cycles. Somehow these configurations produce different manifestations at Earth \-- this has been explored in the context of cosmic rays, but not for other proxies that we are aware of.

\item Slow timescale trends, obtained here using a 40 year (rlowess) local linear regression give a trend in SSN which tracks the Gleissberg cycle. The slow trend in $\langle acv(27) \rangle$,  and indeed in our estimate of the original R27 index, is in antiphase with the SSN trend. These (independent) measures of activity would suggest that the Gleissberg cycle is a very real phenomenon, one rooted in the dynamics controlling the production of the global scale magnetic field, as it is present in both the `dark' and `light' faces of solar activity. We also note that the average sunspot latitude shows a slow trend which tracks the Gleissberg cycle.

\end{itemize}

Our analysis codifies the  Hale cycle-rooted relationship of recurrent solar wind activity and sunspot production. This
provides new insight into why the $aa$-index at solar minimum is a reasonable precursor of the upcoming sunspot cycle amplitude.

\acknowledgments
We thank H. Sargent for illuminating discussions.
SCC acknowledges ST/T000252/1 and FA9550-17-1-0054. SMC is supported by the National Center for Atmospheric Research, which is a major facility sponsored by the National Science Foundation under Cooperative Agreement No. 1852977. SCC, NWW and RJL appreciate the support of the HAO Visitor Program. RJL acknowledges support from NASA's Living With a Star Program. 

\vspace{5mm}

\facilities{All data used in this study is freely available from the following sources:\\
SILSO Royal Observatory of Belgium, Brussels daily total sunspot number version 2.0 from 1818:\\ http://www.sidc.be/silso/home\\
The dates of solar cycle maxima and minima are as determined from the smoothed sunspot number record by SILSO: http://www.sidc.be/silso/cyclesmm\\
Solar radio flux at 10.7 cm (the F10.7 index) from 1947:\\
https://www.spaceweather.gc.ca/solarflux/sx-en.php\\
International Service of Geomagnetic Indices $aa$ index dataset from  1868:\\
http://isgi.unistra.fr/ \\
University of Oulu / Sodankyla Geophysical Observatory GCR flux from 1964:\\
http://cosmicrays.oulu.fi/\\
Royal Observatory, Greenwich - USAF/NOAA Sunspot Data
Sunspot areas from 1874-2016:\\
https://solarscience.msfc.nasa.gov/greenwch.shtml
}




\end{document}